\documentclass[12pt]{iopart}

\usepackage{graphicx} 
\begin{document}

\title[Effects of the Second Harmonic on the GAM in Electron Scale Turbulence]{Effects of the Second Harmonic on the Geodesic Acoustic Mode in Electron Scale Turbulence}

\author{Johan Anderson$^1$, Hans Nordman$^1$ and Raghvendra Singh$^{2,3}$}

\address{$^1$ Dept. Earth and Space Sciences, Chalmers University of Technology, SE-412 96 G\"{o}teborg, Sweden}
\address{$^2$ WCI, Daejon, South Korea} 
\address{$^3$ Institute for Plasma Research, Bhat, Gandhinagar, Gujarat, India 382428}
\ead{anderson.johan@gmail.com}
\begin{abstract}
The effects higher order harmonics have been self-consistently included in the derivation of the electron branch of the electron Geodesic Acoustic Mode (el-GAM) in an Electron-Temperature-Gradient (ETG) turbulence background. The work is based on a two-fluid model including finite $\beta$-effects while retaining non-adiabatic ions. In solving the linear dispersion relation, it is found that the due to the coupling to the $m=2$ mode the real frequency may be significantly altered and yield higher values.  
\end{abstract}

\maketitle

\section{Introduction}
Research during recent years has provided the community with significantly increased  knowledge on the importance of coherent structures such as vortices, streamers and zonal flows ($m=n=0$, where $m$ and $n$ are the poloidal and toroidal mode numbers respectively) in determining the overall transport in magnetically confined plasmas. Zonal flows impede transport by shear decorrelation, whereas the Geodesic Acoustic Mode (GAM) \cite{winsor1968, conway2011, mckee2008, chak2007, miki2007, miki2010, hallatschek2012, diamond2005, terry2000} is the oscillatory counterpart of the zonal flow ($m=n=0$ in the potential perturbation, $m=1$, $n=0$ in the perturbations in density, temperature and parallel velocity) and thus a weaker effect on turbulence is expected. Nevertheless experimental studies suggest that GAMs are related to the L-H transition and transport barriers. The GAMs are weakly damped by Landau resonances and moreover this damping effect is weaker at the edge suggesting that GAMs are more prominent in the region where transport barriers are expected. \cite{mckee2008} Evidence of interactions between the turbulence driven $\vec{E} \times \vec{B}$ zonal flow oscillation or Geodesic Acoustic Mode (GAM), turbulence and the mean equilibrium flows during this transition was found. Furthermore, periodic modulation of flow and turbulence level with the characteristic limit cycle oscillation at the GAM frequency was present. \cite{conway2011} Moreover, in Ref. \cite{waltz2008}, it was observed that GAMs are only somewhat less effective than the residual zonal flow in providing the non-linear saturation. 

For heat transport in the electron channel a likely candidate is the Electron Temperature Gradient (ETG) mode driven by a combination of electron temperature gradients and field line curvature effects. \cite{liu1971, horton1988, jenko2000, singh2001, singh2001NF, tangri2005} The short scale fluctuations that determines the ETG driven heat transport do not influence ion heat transport and is largely unaffected by the large scale flows stabilizing ion-temperature-gradient (ITG) modes. The generation of large scale modes such as zonal flows and GAMs is here realized through the Wave Kinetic Equation (WKE) analysis that is based on the coupling of the micro-scale turbulence with the GAM through the WKE under the assumptions that there is a large separation of scales in space and time. \cite{diamond2005, smol2000, smol2002, krommes2000, anderson2002, anderson2006, anderson2007} In non-linear gyrokinetic simulations large thermal transport levels, beyond mixing length estimates have been observed for a long time. \cite{dorland2000, jenko2000, jenko2002, nevins2006, waltz2008, nakata2012}

In recent work the el-GAM, the finite $\beta$-effects were elaborated on, and numerical quantifications of the frequency and growth rate were given in Refs. \cite{anderson2012, anderson2013}. The finite $\beta$-effects were added in an analogous way compared to the recent work on zonal flows in Ref.~\cite{anderson2011, guzdar2008}. In particular, the Maxwell stress was included in the generation of the el-GAM. The frequency of the el-GAM is higher compared to the ion GAM by the square root of the ion-to-electron mass ratio ($\Omega_q(electron)/\Omega_q(ion) \approx \sqrt{m_i/m_e}$ where $\Omega_q(electron)$ and $\Omega_q(ion)$ are the real frequencies of the electron and ion GAMs, respectively.). It was found that similar to the linear growth rate the finite $\beta$ effects were stabilizing the GAM using a mode coupling saturation level. Furthermore, increasing the non-adiabaticity parameter ($\Lambda_e$) decreased the growth rate through a linear contribution.

It is interesting to note that in simulations, damping of the GAM due to coupling to higher $m$ modes has been found. ~\cite{sugama2006, miyato2007, sasaki2008} In a careful evaluation of the contributions from higher $m$ modes it can be shown that they are, in general, of the order $\epsilon_n$ smaller. However, the effect of higher harmonics is increased by the square of the safety factor ($\bar{q}$) and thus in order to evaluate the effects a more detailed study is called for.

To this end, in this work a detailed investigation of the effects of the higher harmonics on the el-GAM driven by electron temperature gradient (ETG) modes is presented. We have utilized a two-fluid model for the ETG mode based on the Braginskii equations with non-adiabatic ions including impurities and finite $\beta$ - effects. \cite{singh2001, tangri2005} It is shown that the effects of the second harmonics of the density and temperature perturbations on the linear GAM frequency and non-linear generation of the GAM, found in Ref~\cite{anderson2012} can be significant and elevate the frequency of the el-GAM similar to what was discovered in Ref~\cite{elfimov2013}.

The remainder of the paper is organized as follows: In Section II the linear ETG mode including the ion impurity dynamics is presented. The linear el-GAM is presented and the non-linear effects are discussed in Section III. A quantification of the effects of the second harmonics is presented in Sec. IV and the paper is summarized in Sec. V.

\section{The linear Electron Temperature Gradient Mode}
We will start be giving the preliminaries of the Electron-Temperature-Gradient mode described by a two-fluid model. The ETG mode is considered under the following restrictions on real frequency and wavelength: $\Omega_i \leq \omega \sim \omega_{\star} << \Omega_e$, $k_{\perp} c_i > \omega > k_{\parallel} c_e$. Here $\Omega_j$ are the respective cyclotron frequencies, $\rho_j$ the Larmor radii and $c_j = \sqrt{T_j/m_j}$ the thermal velocities. The diamagnetic frequency is $\omega_{\star} \sim k_{\theta} \rho_e c_e / L_n$, $k_{\perp}$ and $k_{\parallel}$ are the perpendicular and the parallel wave numbers. The ETG model consists of a combination of ion and electron fluid dynamics coupled through quasineutrality, including finite $\beta$-effects \cite{singh2001, tangri2005}. First, we will describe the electron dynamics for the toroidal ETG mode governed by the continuity, parallel momentum and energy equations adapted from the Braginskii fluid equations. The electron equations are analogous to the ion fluid equations used for the toroidal ITG mode,
\begin{eqnarray}
\frac{\partial n_{e}}{\partial t} +\nabla \cdot \left( n_{e} \vec{v}_{E} + n_{e} \vec{v}_{\star e} \right) + \nabla \cdot \left( n_{e} \vec{v}_{pe} + n_{e} \vec{v}_{\pi e} \right) + \nabla \cdot \left(n_e \vec{v}_{\parallel e}\right) & = & 0, \label{eq:1.4} \\ 
 \frac{3}{2} n_{e} \frac{\rmd T_{e}}{\rmd t} + n_{e} T_{e} \nabla \cdot \vec{v}_{e} + \nabla \cdot \vec{q}_{e} & = & 0. \label{eq:1.5}
\end{eqnarray}
Here we used the definitions $\vec{q}_e = - (5 p_e/2m_e \Omega_e) \hat{e}_{\parallel} \times \nabla T_e$ as the diamagnetic heat flux, $\vec{v}_{E}$ is the $\vec{E} \times \vec{B}$ drift, $\vec{v}_{\star e}$ is the electron diamagnetic drift velocity, $\vec{v}_{Pe}$ is the electron polarization drift velocity, $\vec{v}_{\pi}$ is the stress tensor drift velocity, and the derivative is defined as $d\rm/\rmd t = \partial/\partial t + \rho_e c_e \hat{e}_{\parallel} \times \nabla \widetilde{\phi} \cdot \nabla$. A relation between the parallel current density and the parallel component of the vector potential ($A_{\parallel}$) can be found using Amp\`{e}re's law,
\begin{eqnarray}\label{eq:1.6}
\nabla^2_{\perp} \widetilde{A}_{\parallel} = - \frac{4 \pi}{c} \widetilde{J}_{\parallel}.
\end{eqnarray}
Taking into account the diamagnetic cancellations in the continuity and energy equations, the Eqs.~(\ref{eq:1.4}, \ref{eq:1.5} and \ref{eq:1.6}) can be simplified and written in normalized form as
\begin{eqnarray}
- \frac{\partial \widetilde{n}_e}{\partial t}  - \nabla_{\perp}^2 \frac{\partial}{\partial t} \widetilde{\phi} - \left( 1  + \left(1 + \eta_e\right) \nabla_{\perp}^2\right) \frac{1}{r} \frac{\partial}{\partial \theta} \widetilde{\phi} - \nabla_{\parallel} \nabla_{\perp}^2 \widetilde{A}_{\parallel}  & + & \nonumber  \\ \epsilon_n \left( \cos \theta \frac{1}{r}\frac{\partial}{\partial \theta} + \sin \theta \frac{\partial}{\partial r} \right)\left(\widetilde{\phi} - \widetilde{n}_e - \widetilde{T}_e\right) & = & \nonumber \\  
-\left(\beta_e/2\right) \left[\widetilde{A}_{\parallel},\nabla_{\parallel}^2 \widetilde{A}_{\parallel}\right] + \left[ \widetilde{\phi}, \nabla^2 \widetilde{\phi}\right], & & \label{eq:1.101} \\
\left(\left(\beta_e/2 - \nabla_{\perp}^2\right) \frac{\partial}{\partial t} + \left(1+\eta_e\right)\left(\beta_e/2\right)\nabla_y\right)\widetilde{A}_{\parallel} + \nabla_{\parallel} \left(\widetilde{\phi} - \widetilde{n}_e - \widetilde{T}_e\right) & = & \nonumber \\
- \left(\beta_e/2\right) \left[\widetilde{\phi} - \widetilde{n}_e, \widetilde{A}_{\parallel}\right]
+ \left(\beta_e/2\right)\left[\widetilde{T}_e,\widetilde{A}_{\parallel}\right] + \left[\widetilde{\phi}, \nabla_{\perp}^2 \widetilde{A}_{\parallel}\right], \label{eq:1.102} \\
\frac{\partial}{\partial t}\widetilde{T}_e + \frac{5}{3} \epsilon_n \left( \cos \theta \frac{1}{r}\frac{\partial}{\partial \theta} + \sin \theta \frac{\partial}{\partial r} \right) \frac{1}{r}\frac{\partial}{\partial \theta} \widetilde{T}_e + \left(\eta_e - \frac{2}{3} \right) \frac{1}{r}\frac{\partial}{\partial \theta} \widetilde{\phi} - \frac{2}{3} \frac{\partial}{\partial t} \widetilde{n}_e & = & -\left[\widetilde{\phi},\widetilde{T}_e\right]. \label{eq:1.103} \nonumber \\
\end{eqnarray}
Note that similar equations have been used previously in estimating the zonal flow generation in ETG turbulence and have been shown to give good agreement with linear gyrokinetic calculations \cite{singh2001, tangri2005}. The variables are normalized according to
\begin{eqnarray}
\left(\widetilde{\phi}, \widetilde{n}, \widetilde{T}_e\right) & = & \left(L_n/\rho_e\right)\left(e \delta \phi/T_{eo}, \delta n_e/n_0, \delta T_e/T_{e0}\right), \\ \label{eq:1.11}
\widetilde{A}_{\parallel} & = & \left(2 c_e L_n/\beta_e c \rho_e\right) e A_{\parallel}/T_{e0}, \\ \label{eq:1.12}
\beta_e & = &  8 \pi n T_e/B_0^2,  \\ \label{eq:1.13}
\epsilon_n & = & \frac{2 L_n}{R}, \\ \label{eq:1.14}
\eta_e & = & \frac{L_n}{L_{T_e}}. \label{eq:1.15}
\end{eqnarray}
Here, $R$ is the major radius and $[A,B] = \frac{\partial A}{\partial r} \frac{1}{r}\frac{\partial B}{\partial \theta} - \frac{1}{r}\frac{\partial A}{\partial \theta} \frac{\partial B}{\partial r}$ is the Poisson bracket. The gradient scale length is defined as $L_f = - (d \ln f/dr)^{-1}$.

Next, we will describe the ion fluid dynamics in the ETG mode description. In the limit $\omega > k_{\parallel} c_e$ the ions are stationary along the mean magnetic field $\vec{B}$ (where $\vec{B} = B_0 \hat{e}_{\parallel}$) whereas in the limit $k_{\perp} c_i >> \omega$, $k_{\perp} \rho_i >> 1$ the ions are unmagnetized. In this paper we will use the non-adabatic responses in the limits $\omega < k_{\perp} c_I < k_{\perp} c_i$, where $c_I = \sqrt{\frac{T_I}{m_I}}$ is the impurity thermal velocity, and we assume that $\Omega_i < \omega < \Omega_e$ are fulfilled for the ions and impurities. In the ETG mode description we can utilize the ion and impurity continuity and momentum equations of the form
\begin{eqnarray} \label{eq:1.1}
\frac{\partial n_j}{\partial t} + n_j \nabla \cdot \vec{v}_j & = & 0, \ \ \mbox{and} \\
m_j n_j \frac{\partial \vec{v}_j}{\partial t} + e n_j \nabla \phi + T_j \nabla n_j & = & 0,
\end{eqnarray}
where $j=i$ for ions and $j=I$ for impurities. Now, we derive the non-adiabatic ion response
with $\tau_i = T_e/T_i$ and impurity response with with $\tau_I = T_e/T_I$, respectively. We thus have
\begin{eqnarray}
\widetilde{n}_j = - \left( \frac{z \tau_j}{1 - \omega^2/\left(k_{\perp}^2 c_j^2\right)}\right) \widetilde{\phi}. \label{eq:1.2}
\end{eqnarray}
Here $T_j$ and $n_j$ are the mean temperature and density of species ($j=e,i,I$), where $\widetilde{n}_i = \delta n/ n_i$, $\widetilde{n}_I = \delta n_I/ n_I$ and $\widetilde{\phi} =  e \phi/T_e$ are the normalized ion density, impurity density and potential fluctuations and $z$ is the charge number of species $j$. Next we present the linear dispersion relation. Using the Poisson equation in combination with (\ref{eq:1.2}) we then find
\begin{eqnarray}
\widetilde{n}_e = - \left( \frac{\tau_i n_i/n_e}{1 - \omega^2/k_{\perp}^2 c_i^2} + \frac{\left(Z^2 n_I/n_e\right) \tau_I}{1 - \omega^2/\left(k_{\perp}^2 c_I^2\right)} + k_{\perp}^2 \lambda_{De}^2\right) \widetilde{\phi}. \label{eq:1.3}
\end{eqnarray}
Considering the linear dynamical equations (\ref{eq:1.101}, \ref{eq:1.102} and \ref{eq:1.103}) and utilizing Eq.~(\ref{eq:1.3}) as in Ref.~\cite{tangri2005} we find a semi-local dispersion relation as follows,
\begin{eqnarray} \label{eq:1.16}
\left[ \omega^2 \left(  \Lambda_e  + \frac{\beta_e }{2} (1 + \Lambda_e ) \right) + \left( 1 - \bar{\epsilon}_n (1 + \Lambda_e) \right) \omega_{\star} \right. & + & \nonumber \\
\left. k_{\perp}^2 \rho_e^2 \left(  \omega - (1 + \eta_e) \omega_{\star} \right) \right] \left( \omega - \frac{5}{3} \bar{\epsilon}_n \omega_{\star} \right) & + & \nonumber \\
\left( \bar{\epsilon}_n \omega_{\star} - \frac{\beta_e}{2} \omega\right) \left( (\eta_e - \frac{2}{3})\omega_{\star} + \frac{2}{3} \omega \Lambda_e \right) & = & \nonumber \\
c_e^2k_{\parallel}^2 k_{\perp}^2 \rho_e^2 \left( \frac{(1 + \Lambda_e) \left( \omega - \frac{5}{3}\bar{\epsilon}_n \omega_{\star}\right) - \left( \eta_e - \frac{2}{3}\right)\omega_{\star} - \frac{2}{3} \omega \Lambda_e}{\omega \left( \frac{\beta_e}{2} + k_{\perp}^2 \rho_e^2\right) - \frac{\beta_e}{2} \left( 1 + \eta_e \right) \omega_{\star}} \right).
\end{eqnarray}
In the following we will use the notation $\Lambda_e = \tau_i (n_i/n_e)/(1 - \omega^2/k_{\perp}^2 c_i^2) + \tau_I (z_\mathrm{eff} n_I/n_e)/(1 - \omega^2/k_{\perp}^2 c_I^2) + k_{\perp}^2 \lambda_{De}^2$. Here we define $z_\mathrm{eff} \approx  z^2 n_I/n_e$. Note that in the limit $T_i = T_e$, $\omega<k_{\perp} c_i$, $k_{\perp} \lambda_{De} < k_{\perp} \rho_e \leq 1$ and in the absence of impurity ions, $\Lambda_e \approx 1$ and the ions follow the Boltzmann relation in the standard ETG mode dynamics. Here $\lambda_{De} = \sqrt{T_c/(4 \pi n_e e^2)}$ is the Debye length, the Debye shielding effect is important for $\lambda_{De}/\rho_e > 1$. The dispersion relation Eq.~(\ref{eq:1.16}) is analogous to the toroidal ion-temperature-gradient mode dispersion relation except that the ion quantities are exchanged to their electron counterparts. Eq.~(\ref{eq:1.16}) is derived by using the ballooning mode transform equations for the wave number and the curvature operator,
\begin{eqnarray}
\nabla_{\perp}^2 \widetilde{f} & = & - k_{\perp}^2 \widetilde{f} = - k_{\theta}^2 \left(1 + \left(s \theta - \alpha \sin \theta\right)^2 \right) \widetilde{f}, \\
\nabla_{\parallel} \widetilde{f} & = & \rmi k_{\parallel} \widetilde{f} \approx \frac{1}{qR} \frac{\partial \widetilde{f}}{\partial \theta}, \\
\widetilde{\epsilon}_n \widetilde{f} & = & \epsilon_n \left( \cos \theta + \left(s \theta - \alpha \sin \theta\right) \sin \theta\right) \widetilde{f} = \epsilon_n g(\theta) \widetilde{f}.
\end{eqnarray} 
The geometrical quantities will be determined using a semi-local analysis by assuming an approximate eigenfunction while averaging the geometry dependent quantities along the field line. The form of the eigenfunction is assumed to be
\begin{eqnarray}\label{eq:1.17}
\Psi(\theta) = \frac{1}{\sqrt{3 \pi}}(1 + \cos \theta) \;\;\;\;\; \mbox{with} \;\;\;\;\; |\theta| < \pi.
\end{eqnarray}
In the dispersion relation we will replace $k_{\parallel} = \left< k_{\parallel} \right>$, $k_{\perp} = \left< k_{\perp} \right>$ and $\omega_D = \left< \omega_D \right>$ by the averages defined through the integrals
\begin{eqnarray}
\left< k_{\perp}^2 \right> & = & \frac{1}{N\left(\Psi\right)}\int_{-\pi}^{\pi} d \theta \Psi k_{\perp}^2 \Psi = k_{\theta}^2 \left( 1 + \frac{s^2}{3} \left(\pi^2 - 7.5\right) - \frac{10}{9} s \alpha + \frac{5}{12} \alpha^2 \right), \label{eq:1.18} \\
\left< k_{\parallel}^2 \right> & = & \frac{1}{N\left(\Psi\right)} \int_{-\pi}^{\pi} d \theta \Psi k_{\parallel}^2 \Psi = \frac{1}{3 q^2 R^2}, \label{eq:1.19}\\
\left< \omega_D \right> & = & \frac{1}{N\left(\Psi\right)} \int_{-\pi}^{\pi} d \theta \Psi \omega_D \Psi = \epsilon_n \omega_{\star} \left( \frac{2}{3} + \frac{5}{9}s - \frac{5}{12} \alpha \right) = \epsilon_n g \omega_{\star},  \label{eq:1.20} \\
\left< k_{\parallel} k_{\perp}^2 k_{\parallel} \right> & = & \frac{1}{N\left(\Psi\right)} \int_{-\pi}^{\pi} d \theta \Psi k_{\parallel} k_{\perp}^2 k_{\parallel} \Psi = \frac{k_{\theta}^2}{3 \left(qR\right)^2} \left( 1 + s^2 \left( \frac{\pi^2}{3} - 0.5 \right) - \frac{8}{3}s \alpha + \frac{3}{4} \alpha^2 \right), \label{eq:1.21} \nonumber \\
\\
N(\Psi) & = & \int_{-\pi}^{\pi} d \theta \Psi^2. \label{eq:1.22}
\end{eqnarray}
Here we have from the equilibrium $\alpha = \beta q^2 R \left(1 + \eta_e + (1 + \eta_i) \right)/(2 L_n)$ and $\beta = 8 \pi n_o (T_e + T_i)/B^2$ is the plasma $\beta$, $q$ is the safety factor and $s = r q^{\prime}/q$ is the magnetic shear. The $\alpha$-dependent term above (in Eq.\ref{eq:1.16}) represents the effects of Shafranov shift.

\section{Modeling Electron Geodesic Acoustic modes}
In this section we will describe the derivation of the dispersion relation for the electron Geodesic Acoustic Modes including the $m=2$ higher harmonic coupling to the $m=1$ and $m=0$ components. The GA mode is defined as having $m=n=0$, $k_r \neq 0$ perturbation of the potential field and the $n=0$, $m=1$, $k_r \neq 0$ perturbation in the density, temperatures and the magnetic field perturbations.~\cite{winsor1968, diamond2005} In addition we will now consider the $m=2$ components of the density, temperature and magnetic field perturbations. The GAM ($q, \Omega_q$) induced by ETG modes ($k,\omega$) is considered under the conditions when the ETG mode real frequency satisfies $\Omega_e > \omega > \Omega_i$ at the scale $k_{\perp } \rho_e < 1$ and the real frequency of the GAM fulfils $\Omega_q \sim c_e/R$ at the scale $q_r < k_{r}$. We start by deriving the linear electron GAM dispersion relation following the outline in the previous paper Ref.~\cite{anderson2012, anderson2013}, by writing the $m=1$ and $m=2$ equations for the density, parallel component of the vector potential and temperature, and the $m=0$ of the electrostatic potential, respectively. Starting with the $m=0$ component,
\begin{eqnarray} 
- \nabla_{\perp}^2 \frac{\partial}{\partial t} \widetilde{\phi}^{(0)}_G - \epsilon_n \sin \theta \frac{\partial}{\partial r} \left( \widetilde{n}^{(1)}_{eG} + \widetilde{T}^{(1)}_{eG} \right) & = & 0, \label{eq:2.1}
\end{eqnarray}
and then the $m=1$ Equations, 
\begin{eqnarray} 
- \frac{\partial \widetilde{n}^{(1)}_{eG}}{\partial t} + \epsilon_n \sin \theta \frac{\partial}{\partial r} \left( \widetilde{\phi}^{(0)}_G - \left( \widetilde{n}^{(2)}_{eG} + \widetilde{T}^{(2)}_{eG} \right) \right) - \nabla_{\parallel} \nabla_{\perp}^2 \widetilde{A}_{\parallel G}^{(1)} & = & 0, \label{eq:2.2} \\ 
\left(\beta_e/2 - \nabla_{\perp}^2\right) \frac{\partial}{\partial t}\widetilde{A}^{(1)}_{\parallel G} - \nabla_{\parallel} \left( \widetilde{n}^{(1)}_{eG} + \widetilde{T}^{(1)}_{eG} \right) & = & 0, \label{eq:2.3} \\
\frac{\partial}{\partial t}\widetilde{T}^{(1)}_{eG} - \frac{2}{3} \frac{\partial}{\partial t} \widetilde{n}^{(1)}_{eG} & = & 0. \label{eq:2.4}
\end{eqnarray} 
Finally the $m=2$ Equations,
\begin{eqnarray} 
-\frac{\partial \widetilde{n}_{eG}^{(2)}}{\partial t} - \nabla_{\parallel} \nabla_{\perp}^{2} \widetilde{A}_{\parallel G}^{(2)} - \epsilon_n \sin \theta \frac{\partial}{\partial r} \left(\widetilde{n}^{(1)}_{eG} + \widetilde{T}^{(1)}_{eG} \right) & = & 0, \label{eq:2.5} \\
\left( \frac{\beta_e}{2} - \nabla_{\perp}^{2} \right) \frac{\partial}{\partial t} \widetilde{A}_{\parallel G}^{(2)} - \nabla_{\parallel} \left( \widetilde{n}^{(2)}_{eG} + \widetilde{T}^{(2)}_{eG}\right) & = & 0, \label{eq:2.6} \\
\frac{\partial}{\partial t} \widetilde{T}^{(2)}_{eG} + \frac{5}{3} \epsilon_n \sin \theta \frac{\partial}{\partial r} \widetilde{T}^{(1)}_{eG} - \frac{2}{3} \frac{\partial}{\partial t} \widetilde{n}^{(2)}_{eG} & = & 0. \label{eq:2.7}
\end{eqnarray} 
Using Eqs. (\ref{eq:2.1}) - (\ref{eq:2.7}), we will derive the linear GAM frequency, by obtaining a relation of the form $\widetilde{T}^{(1)}_{eG} = C_0 \widetilde{n}^{(1)}_{eG}$ eliminating the $m=2$ components. We continue by noting that the Eqs. (\ref{eq:2.4}) and (\ref{eq:2.7}) are symmetric in $m$ using the Fourier representation we find,
\begin{eqnarray}
\widetilde{T}^{(1)}_{eG} & = & \frac{2}{3} \widetilde{n}^{(1)}_{eG} + \frac{5}{3} \frac{\epsilon_n q_r}{\Omega_q} \sin \theta \  \widetilde{T}^{(2)}_{eG}, \label{eq:2.8} \\
\widetilde{T}^{(2)}_{eG} & = & \frac{2}{3} \widetilde{n}^{(2)}_{eG} + \frac{5}{3} \frac{\epsilon_n q_r}{\Omega_q} \sin \theta \ \widetilde{T}^{(1)}_{eG}. \label{eq:2.9}
\end{eqnarray}
We will use Eq. (\ref{eq:2.8}) to derive a relation between the second harmonic ($m=2$) of the density perturbation expressed in terms of the first harmonic ($m=1$) variables. Eqs (\ref{eq:2.5}) and (\ref{eq:2.6}) yield,
\begin{eqnarray}
-\Omega_q \widetilde{n}^{(2)}_{eG} + q_{\parallel} q_{\perp}^2 \widetilde{A}^{(2)}_{\parallel G} - \epsilon_n \sin \theta \ q_r (\widetilde{n}^{(1)}_{eG} + \widetilde{T}^{(1)}_{eG}) & = & 0, \label{eq:2.10} \\
\left( \frac{\beta_e}{2} + q_{\perp}^2 \right) \Omega_q \widetilde{A}^{(2)}_{\parallel G} + q_{\parallel} \left( \widetilde{n}^{(2)}_{eG} + \widetilde{T}^{(2)}_{eG} \right) & = & 0. \label{eq:2.11}
\end{eqnarray}
In order to obtain the desired result we use Eq. (\ref{eq:2.11}) and substitute the $m=2$ temperature perturbation by Eq. (\ref{eq:2.9}) and we find,
\begin{equation}
\widetilde{A}^{(2)}_{\parallel G} = - \frac{5}{3}\frac{q_{\parallel}}{\left( \frac{\beta_e}{2} + q_{\perp}^2\right) \Omega_q} \left( \widetilde{n}^{(2)}_{eG} + \frac{\epsilon_n q_r}{\Omega_q} \sin \theta \ \widetilde{T}^{(2)}_{eG} \right). \label{eq:2.12}
\end{equation} 
Now employ Eq. (\ref{eq:2.10}) and eliminate the parallel vector potential finding,
\begin{eqnarray}
\widetilde{n}^{(2)}_{eG} & = & - \frac{5}{3} \frac{q_{\parallel}^2 q_{\perp}^2}{\Omega_q \left( \frac{\beta_e}{2} + q_{\perp}^2\right)} \left(\widetilde{n}^{(2)}_{eG} + \frac{\epsilon_n q_r}{\Omega_q} \sin \theta \ \widetilde{T}^{(1)}_{eG} \right) \\
& - & \frac{\epsilon_n q_r}{\Omega_q} \sin \theta \ \left( \widetilde{n}^{(2)}_{eG} + \widetilde{T}^{(1)}_{eG} \right). \label{eq:2.13}
\end{eqnarray}
Collecting terms and re-arranging, we find a remarkably simple relation for the second harmonic density perturbation in terms of the $m=1$ components, 
\begin{eqnarray}
\widetilde{n}^{(2)}_{eG} & = & - \frac{\epsilon_n q_r}{\Omega_q} \sin \theta \ \left( \frac{1}{C_1} \widetilde{n}^{(1)}_{eG} + \widetilde{T}^{(1)}_{eG} \right), \label{eq:2.14}\\
C_1 & = & 1 + \frac{5}{3} \frac{q_{\parallel}^2 q_{\perp}^2}{\Omega_q^2 \left( \frac{\beta_e}{2} + q_{\perp}^2 \right)}. \label{eq:2.15}
\end{eqnarray}
Now the relation between the $m=1$ components of temperature and density will be determined by using Eqs. (\ref{eq:2.8}) and (\ref{eq:2.9}),
\begin{eqnarray}
\widetilde{T}^{(1)}_{eG} & = & \frac{2}{3} \widetilde{n}^{(1)}_{eG} + \frac{10}{9} \frac{\epsilon_n q_r}{\Omega_q} \sin \theta \ \widetilde{n}^{(2)}_{eG} \nonumber \\
& + & \frac{25}{9} \frac{\epsilon_n^2 q_r^2}{\Omega_q^2} \sin^2\theta \ \widetilde{T}^{(1)}_{eG}. \label{eq:2.16}
\end{eqnarray}
Collecting terms and eliminating the $m=2$ density perturbation gives,
\begin{eqnarray}
\widetilde{T}^{(1)}_{eG} & = & \frac{2}{3} \frac{1 - \frac{5}{3}\frac{\epsilon_n^2 q_r^2}{\Omega_q^2 C_1} \sin^2 \theta}{1 - \frac{5}{3}\frac{\epsilon_n^2 q_r^2}{\Omega_q^2} \sin^2 \theta} \widetilde{n}^{(1)}_{eG}, \label{eq:2.17} \\
C_0 & = & \frac{2}{3} \frac{1 - \frac{5}{3}\frac{\epsilon_n^2 q_r^2}{\Omega_q^2 C_1} \sin^2 \theta}{1 - \frac{5}{3}\frac{\epsilon_n^2 q_r^2}{\Omega_q^2} \sin^2 \theta}. \label{eq:2.18}
\end{eqnarray}
We have now obtained the desired coefficient $C_0$. Note that, neglecting contributions from the $m=2$ couplings $C_1 = 1$ and the previous relation between the density and temperature is recovered. A key element in determining the dispersion relation, is the relation between the $\widetilde{\phi}_G^{(0)}$ and the $m=1$ density perturbation $\widetilde{n}^{(1)}_{eG}$, this is found in similar way as in Refs. \cite{anderson2012, anderson2013} by using Eq. (\ref{eq:2.2}) as,
\begin{eqnarray}
& \widetilde{n}^{(1)}_{eG} & \left[ 1 - \frac{q_{\parallel}^2 q_{\perp}^2}{\Omega_q^2} \frac{1+C_0}{\frac{\beta_e}{2} + q_{\perp}^2} + \frac{5}{3} \frac{\epsilon_n^2 q_r^2}{\Omega_q^2 C_1} \sin^2 \theta \right] \nonumber \\
& - & \frac{\epsilon_n q_r}{\Omega_q} \sin \theta \ \widetilde{\phi}_G^{(0)} = 0, \label{eq:2.19}
\end{eqnarray}
while noting that there is a simple relation for $\widetilde{n}^{(2)}_{eG} + \widetilde{T}^{(2)}_{eG}$ as,
\begin{equation}
\widetilde{n}^{(2)}_{eG} + \widetilde{T}^{(2)}_{eG} = - \frac{5}{3} \frac{\epsilon_n q_r}{\Omega_q C_1} \sin \theta \ \widetilde{n}^{(1)}_{eG}. \label{eq:2.20}
\end{equation}
We can now determine the dispersion relation for the GAM by considering the $m=0$ component in Eq. (\ref{eq:2.1}) and in addition employ Eqs. (\ref{eq:2.3}) and (\ref{eq:2.19}),
\begin{eqnarray}
\left[ 1 - \frac{q_{\parallel}^2 q_{\perp}^2}{\Omega_q^2} \frac{1+C_0}{\frac{\beta_e}{2} + q_{\perp}^2} + \frac{5}{3} \frac{\epsilon_n^2 q_r^2}{\Omega_q^2 C_1} \sin^2 \theta \right] = \frac{q_r^2}{q_{\perp}^2} \frac{\epsilon_n^2}{\Omega_q^2}\left(1 + C_0 \right) \sin^2 \theta. \label{eq:2.21}
\end{eqnarray}
Here, we employ averaging of the sine components as $\left< \sin^2 \theta \right> = 1/2$ over the poloidal angle $\theta$. We note that neglecting the $m=2$ contributions the coefficient $C_0 = \frac{2}{3}$ and $C_1 = 1$. Note that, in the limit of vanishing temperature perturbations $C_0$ would be zero. Furthermore, the third term on the right hand side comes from the coupling to the $m=2$ component. This is to be compared to the regular GAM frequency found in Refs. \cite{anderson2012, anderson2013},
\begin{eqnarray}  \label{eq:3.9}
\Omega_q^2 = \frac{5}{3}\frac{c_e^2}{R^2} \left( 2 + \frac{1}{\bar{q}^2} \frac{1}{1 + \beta_e/\left(2 q_r^2\right)} \right). \label{eq:2.22}
\end{eqnarray}
Here, $\bar{q}$ is the safety factor. Note that the linear electron GAM is purely oscillating analogously to its ion counterpart c.f. Ref.~\cite{chak2007} and its frequency is decreasing with increasing $\bar{q}$. Here it is of interest to note that it is very similar to the result found in Ref. ~\cite{chak2012}. In order for the GAM to be unstable a non-linear driving by the ETG background is needed. The non-linear state was presented in detail in Refs. \cite{anderson2012, anderson2013} and thus only the main result is given. The non-linear extension to the evolution equations presented previously in Eqs.~(\ref{eq:1.101})--(\ref{eq:1.103}) are
\begin{eqnarray} 
- \frac{\partial \widetilde{n}_e}{\partial t}  - \nabla_{\perp}^2 \frac{\partial}{\partial t} \widetilde{\phi} - \left( 1  + \left(1 + \eta_e\right) \nabla_{\perp}^2\right)\nabla_{\theta} \widetilde{\phi} - \nabla_{\parallel} \nabla_{\perp}^2 \widetilde{A}_{\parallel}  & + & \nonumber  \\ \epsilon_n \left( \cos \theta \frac{1}{r}\frac{\partial}{\partial \theta} + \sin \theta \frac{\partial}{\partial r} \right)\left(\widetilde{\phi} - \widetilde{n}_e - \widetilde{T}_e\right) & = &  \nonumber \\ 
+ \left[ \widetilde{\phi}, \nabla^2 \widetilde{\phi}\right] - \left(\beta_e/2\right) \left[\widetilde{A}_{\parallel}, \nabla^2 \widetilde{A}_{\parallel}\right], \label{eq:7.101} \\ 
\left(\left(\beta_e/2 - \nabla_{\perp}^2\right) \frac{\partial}{\partial t} + \left(1+\eta_e\right)\left(\beta_e/2\right)\nabla_{\theta}\right)\widetilde{A}_{\parallel} + \nabla_{\parallel} \left(\widetilde{\phi} - \widetilde{n}_e - \widetilde{T}_e\right) & = & \left[\widetilde{\phi}, \nabla_{\perp}^2 \widetilde{A}_{\parallel}\right], \label{eq:7.102} \nonumber \\ \\
\frac{\partial}{\partial t}\widetilde{T}_e + \frac{5}{3} \epsilon_n \left( \cos \theta \frac{1}{r}\frac{\partial}{\partial \theta} + \sin \theta \frac{\partial}{\partial r} \right) \frac{1}{r}\frac{\partial}{\partial \theta} \widetilde{T}_e + \left( \eta_e - \frac{2}{3} \right) \frac{1}{r}\frac{\partial}{\partial \theta} \widetilde{\phi} - \frac{2}{3} \frac{\partial}{\partial t} \widetilde{n}_e & = & -\left[\widetilde{\phi},\widetilde{T}_e\right]. \label{eq:7.103} \nonumber \\ 
\end{eqnarray}
Here we will keep the non-linear term in the $m=0$ component whereas all the others can be considered small due to the fact that in evaluating the non-linear terms a summation over the spectrum is performed and that the $m=1$ non-linear terms are odd and thus yield a negligible contribution to the non-linear generation of the GAM. The non-linear contribution to the  $m=0$ potential perturbations are,
\begin{eqnarray} \label{eq:3.2}
- \nabla_{\perp}^2 \frac{\partial}{\partial t} \widetilde{\phi}^{(0)}_G - \epsilon_n \sin \theta \frac{\partial}{\partial r} \left( \widetilde{n}^{(1)}_{eG} + \widetilde{T}^{(1)}_{eG} \right) & = & \nonumber \\
 \left< \left[\widetilde{\phi}_k, \nabla^2 \widetilde{\phi}_k\right] \right>^{(0)} - \left(\beta_e/2\right) \left< \left[\widetilde{A}_{\parallel k}, \nabla^2 \widetilde{A}_{\parallel k}\right] \right>^{(0)} & = & N_2^{(0)}.
\end{eqnarray}
In order to evaluate the Maxwell stress part in Eq.~(\ref{eq:3.2}), we will approximate the parallel part of the electromagnetic vector potential with the electrostatic potential through a linear relation. The relation $\widetilde{A}_{k \parallel} = A_0 \widetilde{\phi}_k$ is found by using the Eqs.~(\ref{eq:1.102}), (\ref{eq:1.103}) and the non-adiabatic response Eq.~(\ref{eq:1.3}) giving an approximation of the total stress of the form
\begin{eqnarray} \label{eq:3.222}
N_2^{(0)} = (1 - |\Omega_{\alpha}|^2)\left< \left[\widetilde{\phi}_k, \nabla^2 \widetilde{\phi}_k\right] \right>^{(0)}.
\end{eqnarray}
The $\Omega_{\alpha}$ factor is found by using Eq.~(\ref{eq:1.102})
\begin{eqnarray} \label{eq:3.223}
\left|\Omega_{\alpha}\right|^2 = \frac{\beta_e}{2}\left| \frac{k_{\parallel} (1 + \Lambda_e + \varphi_0)}{(\beta_e/2 + k^2_{\perp})\omega - (1 + \eta_e)\beta_e k_{\theta}/2}\right|^2,
\end{eqnarray}
where $\varphi_0$ is determined by the temperature equation
\begin{eqnarray} \label{eq:3.224}
\widetilde{T}_{ek} = \frac{(\eta_e - 2/3)- 2/3\omega \Lambda_e}{\omega + 5/3 \epsilon_n g k_{\theta}} \widetilde{\phi}_k = \varphi_0 \widetilde{\phi}_k,
\end{eqnarray}
and $\Lambda_e$ is determined by the non-adiabatic response condition. The expression Eq. (~\ref{eq:3.223}) for the magnetic flutter non-linearity is comparable to that found in Ref.~\cite{singh2001NF} except that in Eq. (~\ref{eq:3.223}) the adiabatic response is taken into account. Note that $\Omega_{\alpha}$ vanishes at $\beta_e = 0$. The relevant non-linear terms can be approximated in the following form
\begin{eqnarray} 
\left< \left[\widetilde{\phi}_k, \nabla_{\perp}^2 \widetilde{\phi}_k\right] \right> & \approx & \left(1 - \left|\Omega_{\alpha}\right|^2\right) q_r^2 \sum_k k_r k_{\theta} \frac{\left|\omega_r\right|}{\epsilon_0} \delta N_k\left(\vec{q},\Omega_q\right). \label{eq:4.3}
\end{eqnarray}
In order to determine the non-linear generation of el-GAMs by the ETG modes will use the wave kinetic equation \cite{diamond2005, smol2000, smol2002, krommes2000, chak2007, anderson2002, anderson2006, anderson2007} to describe the background short scale ETG turbulence for $(\Omega_q, q) < (\omega, k)$, where the action density $N_k = E_k/|\omega_r| \approx \epsilon_0 |\phi_k|^2/\omega_r$. Here $\epsilon_0 |\phi_k|^2$, is the total energy in the ETG mode with mode number $k$ where $\epsilon_0 = \tau + k_{\perp}^2 + \frac{\eta_e^2 k_\theta^2}{|\omega|^2}$. We assume that for all GAMs we have $q_r > q_{\theta}$, with the following relation between $\delta N_k$ and $\partial N_{k 0} / \partial k_r$,
\begin{eqnarray} \label{eq:4.6}
\delta N_k = -\rmi q_r^2 k_{\theta} \phi^0_G G(\Omega_q) \frac{\partial N_{0k}}{\partial k_r} + \frac{k_{\theta} q_r \widetilde{T}_{e G}^{(1)} N_{0k}}{\tau_i \sqrt{\eta_e - \eta_{eth}}}, 
\end{eqnarray}
where we have used $\delta \omega_q = \vec{k} \cdot \vec{v}_{Eq}\approx \rmi (k_{\theta} q_r - k_r q_{\theta}) \phi^{(0)}_G$ in the wave kinetic equation and the definition $G(\Omega_q) = \frac{1}{\Omega_q - q_r v_{gr} + \rmi \gamma_k}$. Here the linear instability threshold of the ETG mode is denoted by $\eta_{e th}$ and is determined by numerically solving Eq.~(\ref{eq:1.16}). Using the results from the wave-kinetic treatment we can compute the non-linear contributions to be of the form
\begin{eqnarray}
\left< \left[\widetilde{\phi}, \nabla_{\perp}^2 \widetilde{\phi}\right] \right>^{(0)} & = & -\rmi \left(1 - |\Omega_{\alpha}|^2\right) q_r^4 \sum k_r k_{\theta}^2 \frac{|\omega_r|}{\epsilon_0} G\left(\Omega_q\right) \frac{\partial N_k}{\partial k_r} \widetilde{\phi}_G^{(0)}.  \label{eq:4.61}
\end{eqnarray}
We note that the non-linear contribution is purely complex and thus will solely determine the growth rate of the GAMs. The growth rates will behave in the same manner as found in Ref. \cite{anderson2013}, however the real frequency of the GAM will be modified by the $m=2$ contributions.
\section{Results and Discussion}
Here will quantify the effect of the contributions of the higher harmonics to the real frequency of the el-GAM by numerically solving the dispersion relation found in Eq. (\ref{eq:2.21}) while comparing the results with the corresponding values found by using the Eq. (\ref{eq:2.22}).

\begin{figure}[ht!]
\centering
\includegraphics[width=7cm, height = 6cm]{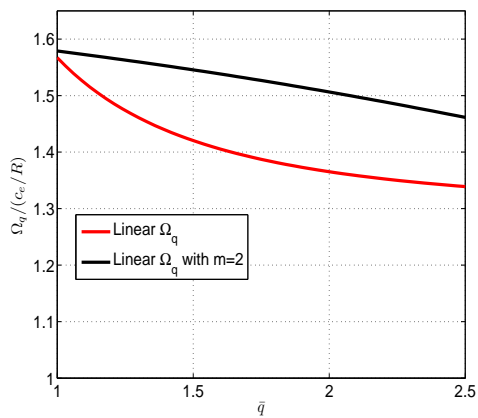}
%\caption{default}
\caption{(color online) The linear el-GAM real frequency (with $m=2$ harmonics included in black line and without represented by the red line) normalized to ($c_e/R$) as a function of the safety factor $\bar{q}$ is shown for the parameter $\eta_e = 4.0$ whereas the remaining parameters are $\epsilon_n = 0.909$, $\beta = 0.01$, $q_x \rho_e = 0.3$ in the strong ballooning limit $g(\theta)=1$.}
\label{fig:fig1}
\end{figure}

In Figure 1, a comparison between the real frequency of the el-GAM with higher harmonics included (black line) and without (red line) is displayed for the parameters $\epsilon_n = 0.909$, $\beta = 0.01$, $q_x \rho_e = 0.3$ in the strong ballooning limit $g(\theta)=1$. The real frequency is decreasing with increasing safety factor $\bar{q}$, according to Eq. (\ref{eq:2.22}). Furthermore, allowing for interactions with the higher harmonics ($m=2$) components moderates the decrease in the frequency. This effect is due to the third term on the left hand side arising from the $m=2$ higher harmonics. Note that, the $C_0$ term describes the effect of including temperature perturbations in the system and would vanish if these could be neglected.

It is expected that the GAM is more prominent for larger values of the safety factor ($\bar{q}$) since it has been showed that for small $\bar{q}$ around unity (core region) GAMs are strongly Landau damped, nevertheless, it seems that including higher harmonics the GAM may attain a higher frequency and that this effect is much stronger for larger $\bar{q}$ (edge region). In Ref. \cite{waltz2008}, it was observed that GAMs are only somewhat less effective than the residual zonal flow in providing the non-linear saturation. In the view of these simulations results, the study here may be of significant importance in the complicated saturation dynamics.  

\section{Summary}
In this work the effects of including higher harmonics ($m=2$) in deriving the dispersion relation for the electron Geodesic Acoustic Mode (el-GAM) are investigated, in previous works this effect have been primarily overlooked. Moreover, it was shown in simulations that coupling to higher order harmonics may significantly influence the dynamics. This work extends previous studies (Ref.~\cite{anderson2012, anderson2013}) by explicitly include the coupling to the $m=2$ harmonic into the el-GAM study. In the model, linear as well as non-linear $\beta$ effects are included in the derivation. The linear dispersion relation of the el-GAM is purely oscillatory with a frequency $\Omega_q \sim \frac{c_e}{R}$ which is decreasing with increasing safety factor ($\bar{q}$). The GAM growth rate is estimated by a non-linear treatment based on the wave-kinetic approach where a competition between the Reynolds stress and the Maxwell stress is present. The linear dispersion relation is solved numerically comparing and quantifying the effect of the coupling to the $m=2$ harmonic. It is found that the decrease in the real frequency of the el-GAM is significantly moderated by the $m=2$ interactions for larger values of the safety factor. However the quantitative results are dependent on the other physical parameters such as the finite $\beta$-effects and the GAM wavevector $q_x$.


\section*{References}
\begin{thebibliography}{100}
%
\bibitem{winsor1968} 
N. Winsor, J. L. Johnson, and J. M. Dawson, Phys. Fluids {\bf 11}, 2448 (1968).
%
\bibitem{conway2011}
G. D. Conway, C. Angioni, F. Ryter, P. Sauter, J. Vicente and the Asdex Upgrade Team \PRL \ {\bf 106}, 065001 (2011).
%
\bibitem{mckee2008} 
G. R. McKee, P. Gohil, D. J. Schlossberg, J. A. Boedo, K. H. Burrell, J. S. deGrassie, R. J. Groebner, R. A. Moyer, C. C. Petty, T. L. Rhodes, L. Schmitz, M. W. Shafer, W. M. Solomon, M. Umansky, G. Wang, A. E. White, and X. Xu, Nuclear Fusion {\bf 49}, 115016 (2009).
%
\bibitem{chak2007} 
N. Chakrabarti, R. Singh, P. Kaw and P. N. Guzdar, Phys. Plasmas {\bf 14}, 052308 (2007).
%
\bibitem{miki2007} 
K. Miki, Y. Kishimoto, N. Miyato and J. Li, Phys. Rev. Lett. {\bf 99}, 145003 (2007)
%
\bibitem{miki2010} 
K. Miki and P. H. Diamond, Phys. Plasmas {\bf 17}, 032309 (2010).
%
\bibitem{hallatschek2012} 
R. Hager and K. Hallatschek, Phys. Rev. Lett. {\bf 108}, 035004 (2012).
%
\bibitem{diamond2005} 
P. H. Diamond, S.-I. Itoh, K. Itoh and T. S. Hahm, \PPCF \ {\bf 47} R35 (2005).
%
\bibitem{terry2000} 
P. W. Terry, Reviews of Modern Physics 72, 109 (2000).
%
\bibitem{waltz2008}
R. Waltz and C. Holland, Phys. Plasmas {\bf 15}, 122503 (2008).
%
\bibitem{liu1971} 
C. S. Liu, Phys. Rev. Lett. {\bf 27}, 1637 (1971).
%
\bibitem{horton1988} 
W. Horton, B. G. Hong and W. M. Tang, Phys. Fluids {\bf 31}, 2971 (1988).
%
\bibitem{jenko2000} 
F. Jenko, W. Dorland, M. Kotschenreuter and B. N. Rogers, Phys. Plasmas {\bf 7}, 1904 (2000).
%
\bibitem{singh2001} 
R. Singh, V. Tangri, H. Nordman and J. Weiland, Phys. Plasmas {\bf 8}, 4340 (2001).
%
\bibitem{singh2001NF}
R. Singh, P. K. Kawand J. Weiland, Nuclear Fusion {\bf 41}, 1219 (2001).
%
\bibitem{tangri2005} 
V. Tangri, R. Singh and P. K. Kaw, Phys. Plasmas {\bf 12}, 072506 (2005).
%
\bibitem{smol2000} 
A. I. Smolyakov, P. H. Diamond and M. V. Medvedev, Phys. Plasmas, {\bf 7} 3987 (2000).
%
\bibitem{smol2002} 
A. I. Smolyakov, P. H. Diamond and Y. Kishimoto, Phys. Plasmas {\bf 9}, 3826 (2002).
%
\bibitem{krommes2000} 
J. A. Krommes and C.-B. Kim, Phys. Rev. E {\bf 62}, 8508 (2000).
%
\bibitem{anderson2002} 
J. Anderson, H. Nordman, R. Singh and J. Weiland, Phys. Plasmas {\bf 9}, 4500 (2002).
%
\bibitem{anderson2006} 
J. Anderson, H. Nordman, R. Singh and J. Weiland, Plasma Phys Controlled Fusion {\bf 48}, 651 (2006).
%
\bibitem{anderson2007} 
J. Anderson and Y. Kishimoto, Phys. Plasmas {\bf 14}, 012308 (2007).
%
\bibitem{dorland2000}
W. Dorland, F. Jenko, M. Kotschenreuter and B. N. Rogers, Phys. Rev. Lett. {\bf 85}, 5579 (2000).
%
\bibitem{jenko2002} 
F. Jenko and W. Dorland, Phys. Rev. Lett. {\bf 89}, 225001 (2002).
%
\bibitem{nevins2006}
W. M. Nevins, J. Candy, S. Cowley, T. Dannert, A. Dimits, W. Dorland, C. Estrada-Mila, G. W. Hammett, F. Jenko, M. J. Pueschel and D. E. Shumaker, Phys. Plasmas {\bf 13}, 122306 (2006).
%
\bibitem{nakata2012}
M. Nakata, T.-H. Watanabe and H. Sugama, {\bf 19}, 022303 (2012).
% 
\bibitem{anderson2012} 
J. Anderson, H. Nordman, R. Singh and P. K. Kaw, Phys. Plasmas {\bf 19}, 082305 (2012).
%
\bibitem{anderson2013}
J. Anderson, A. Skyman, H. Nordman, R. Singh and P. K. Kaw, Nuclear Fusion, in press (2013).
%
\bibitem{anderson2011}
J. Anderson, H. Nordman, R. Singh and R. Singh, Phys. Plasmas {\bf 18}, 072306 (2011).
%
\bibitem{guzdar2008} 
P. N. Guzdar, N. Chakrabarti, R. Singh and P. K. Kaw, Plasma Phys. Contr. Fusion {\bf 50}, 025006 (2008).
%
\bibitem{sugama2006}
H. Sugama and T.-H. Watanabe, J. Plasma Physics {\bf 72}, 825 (2006).
%
\bibitem{miyato2007}
N. Miyato, Y. Kishimoto and J. Li, Nuclear Fusion {\bf 47}, 929 (2007).
%
\bibitem{sasaki2008}
M. Sasaki, K. Itoh, A. Ejiri and Y. Takase, Plasma Fusion Research {\bf 3}, 009 (2008).
%
\bibitem{elfimov2013} 
A. Elfimov, A. Smolyakov, A. Melnikov and R. Galvao, Phys. Plasmas {\bf 20}, 052116 (2013).
%
\bibitem{chak2012}
N. Chakrabarti, P. N. Guzdar and P. K. Kaw, Phys. Plasmas {\bf 19}, 092113 (2012).
%
\end{thebibliography}
\end{document}